\documentclass{article}
\usepackage{amssymb,amsmath}
\usepackage{doublespace}
\usepackage{epsf}

\begin{document}
\begin{spacing}{1.0}

\title{NONCLASSICAL KINETICS IN CONSTRAINED GEOMETRIES: INITIAL
DISTRIBUTION EFFECTS} 

\author{
K. Lindenberg, A. H. Romero and J. \ M. Sancho \footnote{Permanent address:
Departament d'Estructura i Constituents de la
Mat\`{e}ria,
Universitat de Barcelona, 
Av. Diagonal 647, E-08028 Barcelona, Spain.}\\
Department of Chemistry and Biochemistry and Institute for Nonlinear
Science\\
University of California, San Diego,\\
La Jolla, CA 92093-0340, USA}

\date{\today}

\maketitle

\begin{abstract}
We present a detailed study of the effects of the initial distribution
on the kinetic evolution of the irreversible reaction $A+B\rightarrow 0$
in one dimension. Our analytic as well as numerical work is based on a
reaction-diffusion model of this reaction. We focus on the role of initial
density fluctuations in the creation of the macroscopic patterns that
lead to the well-known kinetic anomalies in this system.  In particular,
we discuss the role of the long wavelength components of the initial
fluctuations in determining the long-time
behavior of the system. We note that the frequently
studied random initial distribution is but one of a variety of
possible distributions leading to interesting anomalous behavior.
Our discussion includes an initial distribution with correlated $A-B$
pairs and one in which the initial distribution forms a fractal pattern.
The former is an example of a distribution whose long wavelength components
are suppressed, while the latter exemplifies one whose long wavelength
components are enhanced, relative to those of the random distribution,
\end{abstract}

\thispagestyle{empty}

\section{Introduction}
\label{intro}

Diffusion-limited reactions in low-dimensional geometries
are well known to exhibit \lq\lq anomalous" kinetics different
from those predicted by the standard law of mass action
[Ovchinnikov \& Zeldovich, 1978; Toussaint \& Wilzcek, 1983;
Kang \& Redner, 1984; Kopelman, 1988; Burlatskii {\em et al.}, 1989].
An example of anomalous kinetics is exhibited by the diffusion-limited
irreversible reaction $A+B\rightarrow 0$.  If the distribution of
each reactant is spatially homogeneous,
the rate laws for the global reactant densities $\rho_A(t)$ and
$\rho_B(t)$ are
${\dot \rho}_A = {\dot \rho}_B =-K \rho_A \rho_B$,
where $K$ is
a constant global rate coefficient.  If
$\rho_A(0) = \rho_B (0)\equiv\rho (0)$ then the densities of the two
species are equal at all times, $\rho_A(t) = \rho_B (t)\equiv\rho (t)$, 
and we can write
$\dot\rho=- K \rho^2$. In integrated form
$\rho^{-1}(t)=\rho(0)^{-1} + Kt$; at long times
$\rho\sim (Kt)^{-1}$.  We call this behavior \lq\lq classical."
However, the actual asymptotic rate
law in an infinite volume in dimensions $d<4$ for an
initially random distribution of reactants
is instead $\dot\rho =-K\rho^{(1+4/d)}$ 
[Ovchinnikov \& Zeldovich, 1978; Toussaint \& Wilzcek, 1983;
Kang \& Redner, 1984; Kopelman, 1988; Burlatskii {\em et al.}, 1989].
The associated asymptotic time dependence of the density is then
$\rho\sim t^{-d/4}$ (with logarithmic corrections for $d=4$).

The physical origin of the \lq\lq anomalies" is in general clearly
understood. The usual law of mass action assumes a spatially homogeneous
mixture of reactants; however, in low
dimensions diffusion is not an efficient mixing mechanism and
may be incapable of smoothing out reactant concentration fluctuations
that are created, for example,
by nonhomogeneous initial conditions, by external reactant sources
if the system is open, or by the reaction itself.

Consider the $A+B\rightarrow 0$ reaction starting from
an initially random distribution of $A$ and $B$. Such a distribution
exhibits local density fluctuations, that is, certain
regions of the system are relatively rich in species $A$ while others
are relatively rich in species $B$.  
By \lq\lq efficient" mixing we mean mixing that is rapid relative to the
local reaction rate. Thus, if efficient mixing (by whatever mechanism)
occurs, such local fluctuations are rapidly suppressed.
If mixing is not efficient as, for example, in a diffusion-limited
reaction in low dimensions, then
the reaction will cause the local minority species
to be eliminated extremely rapidly by
the local majority species, and diffusion is not able to homogenize the
system.  There follows an evolution of regions in which
essentially only one or the other of the species is present, and any 
attempt of the minority species to diffuse into this region
leads to its rapid disappearance by reaction.  The reaction
then only takes place at the interfaces of aggregates of different
species.  As time proceeds (and the concentrations of reactants decrease),
the regions in space that are occupied by reactants of essentially
only one species
grow in size, and the total interface area at which the reaction can occur
decreases.  This leads to ever more effective spatial segregation of the
species and an associated slowing down of the reaction relative to its rate
in a homogeneous mixture.  The system goes through
a hierarchy of anomalous kinetic behaviors
[Argyrakis {\em et al.}, 1993; Lindenberg {\em et al.}, 1996]
to arrive asymptotically at the so-called Zeldovich regime
[Ovchinnikov \& Zeldovich, 1978; Burlatskii {\em et al.}, 1989]
characterized by the time dependence $\rho \sim t^{-d/4}$ for $d<4$. 

This discussion punctuates three interesting features of these irreversible
diffusion-limited reactions in constrained (low dimensional)
geometries.  One is that the initial distribution determines
the evolution of the system for all time: the future of the reaction is
imprinted in the spatial distribution at one given (\lq\lq initial")
moment. A second is the fact that it is necessary to start with
{\em disorder} in order to produce an {\em ordered} state at a later time:
the formation of ever larger aggregates of a single species is a direct
consequence of initial fluctuations that are not homogenized. 
A third is the realization that a random distribution is but one of many
possible initial distributions, and it is certainly not the
most easily realized in practice. The historical focus on the random
distribution has perhaps obscured the fact that other initial
distributions lead to other behaviors, some \lq \lq less anomalous"
and others even \lq\lq more anomalous" than the Zeldovich law
$\rho \sim t^{-d/4}$ in their deviation from classical kinetics. 
It is this last point that we focus on in this paper.

It is reasonable to expect that the suppression or
enhancement of initial density fluctuations will
affect the kinetic progression and change even the asymptotic behavior of
the $A+B\rightarrow 0$ system. 
In particular, an initial distribution that is more homogeneous or mixed
than a random one is expected to lead to behavior closer to classical;
conversely, an initial distribution in which fluctuations are
more prominent is expected to lead to greater deviations from the
law of mass action.
Our work expands on previous concern with this problem by
a number of authors
[Vitukhnovsky {\em et al.}, 1988; Abramson {\em et al.}, 1994;
Abramson, 1995; Lindenberg {\em et al.}, 1994, 1996; Sancho {\em et al.},
1996].

A \lq\lq mixing mechanism" that we must mention at the outset
because it necessarily affects the numerical simulations that parallel
our theoretical analysis (and because it occurs in physical systems)
is provided by the boundaries of a {\em
finite} system of volume $V=L^d$, where $L$ is the linear dimension
of the system. Regardless of the precise nature of the boundaries
(reflective, periodic), segregation of species is in any case
limited by the finite size of the system.  With periodic boundaries a
reactant \lq\lq perceives" this finite size at a time
[Lindenberg {\em et al.}, 1996]
\begin{equation}
t_f \sim L^2/8\pi^2D,
\label{finitesizet}
\end{equation}
where $D$ is the diffusion coefficient for that reactant (the numerical
factors are slightly different for other boundary conditions).\footnote
{The numerical factors in (\ref{finitesizet}) arise from the first
non-zero eigenvalue of the diffusion equation.  The time to cross a
distance $L$ on average in a diffusion process would lead to the estimate
$t_f\sim L^2/6D$. In other words, this is only
meant as an order-of-magnitude estimate.}

In Sec.\ \ref{summary} we summarize the
model and our general theoretical framework. 
In Sec.\ \ref{initial} we study various initial conditions 
and present for each case our theoretical and numerical results.
A summary of conclusions is presented in
Sec.\ \ref{conclusions}.

\section{Theoretical Framework}
\label{summary}

We base our analysis on the reaction-diffusion equations 
utilized in many theoretical studies of this problem.  
It is important to point out that our numerical results are also based on
these {\em same} equations.  This differs from the more usual,
more microscopic, approach of simulating
the system as a lattice gas in which
the reactants are random walkers and in which the reaction is an
annihilation that occurs with certainty or with some probability
when reactants step on the same lattice site. 
Simulations at this level of detail
require large computer capabilities and long times to reach
asymptotic behaviors
[Argyrakis {\em et al.}, 1993; Lindenberg {\em et al.}, 1994, 1996].
Reaction-diffusion equations represent mesoscopic approximations to this
more microscopic picture; such approximations
can therefore only provide a theoretical backdrop over
spatial and temporal ranges that are not too detailed and for
densities that are not too low. The results of direct numerical solution
of the reaction-diffusion equations
are of course constrained in the same way as are the equations themselves.
On the other hand, such solutions, being more mesoscopic from
the outset, require less computational intensity and length to
cover the hierarchies of behaviors
and to achieve asymptotic behaviors. Although this approach of course
provides no information on the limits of validity of reaction-diffusion
models, they do provide a simpler and less costly backdrop against which
to check the theoretical analysis of the model.

The reaction-diffusion equations 
for the local densities $\rho_A(\vec{x},t)$ and $\rho_B(\vec{x},t)$ 
of the two species are:

\begin{eqnarray}
\frac{\partial \rho_A}{\partial t} & = & D \nabla^2 \rho_A - K \rho_A
\rho_B
\nonumber\\  
\frac{\partial \rho_B}{\partial t} & = & D \nabla^2 \rho_B - K
\rho_A \rho_B 
\label{difreac} 
\end{eqnarray}
The reaction takes place under stoichiometric conditions for the global
densities,

\begin{equation}
\rho_A(t) \equiv \frac{1}{V}\int d\vec{x} \; \rho_A(\vec{x},t)
= \rho_B(t)\equiv \rho(t),
\label{global}
\end{equation}

\noindent
so that the densities of both species remain equal forever. 
$D$ is the diffusion coefficient, which we assume to be the same
for both species, $K$ is the local reaction rate constant, and $V$
is the system volume.
Note that we are using the same symbol, $K$, for the local rate
coefficient as we did for the global rate coefficient.  When the global
rate law is classical, the two are in fact the same; when it is not, the
global rate is in any case not quadratic in the global density
and the \lq\lq rate coefficient"
in a quadratic rate law is then not even a constant. 
We vary the values of the parameters in our analysis, but recognize that 
(in appropriate dimensionless units) a diffusion-limited reaction occurs
if $D/K \ll 1$ -- indeed, a strictly diffusion-limited reaction requires
$K \rightarrow \infty$.  With $K$ finite (no matter how large), when
the density is sufficiently low the reaction becomes reaction- rather
than diffusion-limited

A few comments concerning notation are useful at this point.  The local
densities $\rho_i(\vec{x},t)$ evolve from initial configurations
$\rho_i(\vec{x},0)$.  In some cases these initial configurations are chosen
from a distribution of different configurations (e.g., a random
distribution) and it might be appropriate to perform an average over the
initial distribution
[Lindenberg {\em et al.}, 1996].
In other cases the initial configuration may be a
very specific one, e.g. as in the study of the evolution of fronts
[Koo \& Kopelman, 1991].
We denote an average of a quantity $f(\vec{x},t)$ over
the initial distribution by
a bracket: $\left< f(\vec{x},t)\right>$.  If we are dealing with a specific
initial configuration then the initial distribution is a delta function
and a quantity \lq\lq averaged over the initial
distribution" is simply the same as the unaveraged quantity,
$\left< f(\vec{x},t)\right> = f(\vec {x},t)$.  

Global quantities such as the global density are
related to local ones by the volume average indicated
in Eq.~(\ref{global}). We also use a single bracket notation to denote
an average over volume.  The use of a single bracket to denote one
or the other or both averages
should not create confusion since the context makes it clear what averages
have been taken.  In any case, a dependence only on the time
indicates that a volume average has been performed,
$f(t)=\left< f(\vec{x},t)\right>$. 

The first impetus toward finding the global density from
the reaction-diffusion equation might be
to average Eq.~(\ref{difreac}) over initial distribution and volume so as
to obtain an equation for the rate of change of the global density.
However, this of course introduces a two-density
average on the right hand side: 
\begin{equation}
\frac{\partial\left<\rho_A(\vec{x},t)\right>}{\partial t} =
\frac{\partial\left<\rho_B(\vec{x},t)\right>}{\partial t} =
- K \left<\rho_A \rho_B\right>,
\label{integrated} 
\end{equation}
where we have noted that $\nabla^2\rho(t)=0$
since $\rho(t)$ is independent of $\vec {x}$.  
One might then write an equation for the two-density
product and again perform an average, which introduces a three-density
term. In many problems it is possible to truncate such a
hierarchy and thereby close the set of equations.  Such closure is not
possible here because it would miss the many-body nature of the problem:
indeed, the evolution of macroscopic structures in these systems
would seem to indicate
that no \lq\lq finite body approximation" is appropriate, and that one
needs to capture at least some aspects of the many-body effects to all
orders to obtain a reasonable approximation.

We note that classical behavior is captured by Eq.~(\ref{integrated})
only if the average of the product can be broken into the product of the
averages, $\left< \rho_A\rho_B\right>=\left< \rho_A\right>
\left<\rho_B\right>$.  This separation is in general not correct and, in
particular, not when there is species segregation.  Indeed, if the species
are strictly segregated then the right hand side is zero because it is not
possible to find $A$ and $B$ at the same location, and thus the rate of
change of the global densities vanishes, as it should.  If there is {\em
some} species overlap, then the decay rate does not vanish but is 
slow (slower than classical).  If the species are randomly or homogeneously
distributed, then the average of the product does reduce to the product of
the averages and the kinetics are classical, with a consequent
inverse time decay of the global densities. 

A number of procedures have been introduced to deal with the many-body
aspects of the problem
[Vitukhnovsky {\em et al.}, 1988; Burlatskii {\em et al.}, 1989].
In our analytic studies we follow [Vitukhnovsky {\em et al.}, 1988;
[Lindenberg {\em et al.}, 1988, 1990; Sokolov \& Blumen, 1991;
Sancho {\em et al.}, 1996]
by introducing the difference and sum variables, 

\begin{equation}
\gamma (\vec{x},t) = \frac{\rho_A(\vec{x},t) - \rho_B(\vec{x},t)}{2}, \qquad
\rho (\vec{x},t) = \frac{\rho_A(\vec{x},t) + \rho_B(\vec{x},t)}{2},
\label{variables}
\end{equation}
so that the pair of equations (\ref{difreac}) transforms into

\begin{equation}
\frac{\partial \gamma}{\partial t} = D \bigtriangledown^2 \gamma 
\label{dif}
\end{equation}

\begin{equation}
\frac{\partial \rho}{\partial t} = D \bigtriangledown^2 \rho - 
K (\rho^2 - \gamma^2).
\label{sum}
\end{equation}
The difference variable $\gamma(\vec{x},t)$ captures the variations
and fluctuations in the spatial distribution of the species.  If the
system is thoroughly mixed, this variable is everywhere small; if the
species are segregated, then the variable is positive in regions where $A$
predominates and negative where $B$ predominates; its variability contains
information on the sizes of such regions.  If the species were oppositely
charged, for example, then this variable would represent the local net
charge.  Fortunately, Eq.~(\ref{dif}) is a simple linear diffusion
equation and can be solved exactly
[Vitukhnovsky {\em et al.}, 1988 Lindenberg {\em et al.}, 1996]
(and it is
through this exact solution that one captures important many-body effects
to all orders):
\begin{equation}
\gamma(\vec{x},t)=\frac{1}{V}\sum_{\vec{k}}\int d\vec{x}'
e^{-Dtk^2}
e^{i\vec{k}\cdot(\vec{x}-\vec{x}')} \gamma(\vec{x}',0)
\label{sln1}
\end{equation}
where $k=|\vec{k}|$.  For periodic
boundary conditions $\vec{k}=2\pi{\vec{n}}/L$ and $\vec{n}$ is a $d$-tuple of
integers.  In the limit $V\rightarrow \infty$ one can transform the sum to
an integral:
\begin{equation}
\gamma(\vec{x},t)=\frac{1}{(2\pi)^d}\int d\vec{k} \int d\vec{x}'
e^{-Dtk^2}
e^{i\vec{k}\cdot(\vec{x}-\vec{x}')} \gamma(\vec{x}',0).
\label{sln2}
\end{equation}
The integration over $\vec k$ can easily be done explicitly in any
dimension; in this paper we deal with one-dimensional systems, 
for which one readily obtains
\begin{equation}
\gamma(x,t)=\frac{1}{2\sqrt{\pi Dt}}\int dx' e^{-(x-x')^2/4Dt}\gamma(x',0),
\label{sln1d}
\end{equation}
where we have dropped the vectorial notation for $x$ and $x'$ but note
that they can be positive or negative.
We return to the difference variable momentarily.

Consider next Eq.~(\ref{sum}) for the sum variable.  This variable is the
density without consideration
of species identity, and it is in general expected to
be a smoother function than the difference variable.  
Let us average Eq.~(\ref{sum})
over the volume and over the initial distribution of reactants:
\begin{equation}
\frac{d\left<\rho\right>}{dt}=-K\left[ \left<\rho^2\right> - \left<
\gamma^2\right> \right].
\label{averagesum}
\end{equation}
The entire subsequent
analysis of the kinetic behavior of the average global density
$\rho(t)\equiv \left<\rho(\vec{x},t)\right>$ 
is based on this equation, and, in particular, on the
behavior of the function
\begin{equation}
G(t)\equiv\left<\gamma^2(\vec{x},t)\right>
\label{driver}
\end{equation}
(which one knows exactly), and on the relation between
$\left<\rho^2(\vec x, t)\right>$ and $\rho(t)$.
We shall call $G(t)$ the segregation function, since the difference
variable (and consequently this function) vanishes for a homogeneous
distribution.

As we demonstrate explicitly below, the decay of $G(t)$
with time (through diffusion) is not simply of
a fixed inverse power form over the entire history of the
process.  Furthermore, in general there is no simple relation
that holds for the entire progression of the reaction
between the average of the square of
the local density and the average density itself.
Consequently, there is
no exact kinetic description of the system in terms of only the global
density: indeed, Eq.~(\ref{averagesum}) is again but the first member
of an infinite moment hierarchy.
Nevertheless, one attempts to extract a kinetic description from
Eq.~(\ref{averagesum}), at least approximately, recognizing that in any
case a single kinetic description does not hold for all times.
It has therefore become common practice to speak of kinetic behaviors
that are valid over some substantial time range, and crossovers from one
type of behavior to another. If the \lq\lq crossover" regimes are short
compared to the regimes where one type of behavior persists, one can then
characterize the system in terms of a sequence of kinetic behaviors and
transitions between these behaviors.  Such a description turns out to be
possible in many cases.  Toward this purpose, in
[Lindenberg {\em et al.}, 1996]
we presented an extensive discussion concerning the
relation between the average of the square of the density,
$\left<\rho^2({\vec x},t)\right>$, and the average density, 
$\rho(t) =\left<\rho({\vec x},t)\right>$.  
In particular, we noted that in
general $\left<\rho ^2(\vec x, t)\right> \neq \rho^2(t)$.
For a random distribution of reactants or of reactant pairs
\begin{equation}
\left< \rho^2(\vec x, t)\right> = \rho^2(t) + \rho_{max}\rho(t),
\label{random}
\end{equation}
where $\rho_{max}$ is the \lq\lq maximum density"
[that is, $\rho(t) \leq \rho_{max}$] specified later for our
particular reaction-diffusion approach.\footnote{From a molecular
viewpoint $\rho_{max}$ is the inverse
of the volume
of one molecule.  In a lattice approach it is the maximum number of walkers
that can simultaneously occupy each lattice site.}
On the other hand, for a segregated distribution as occurs in the Zeldovich
regime, and also for a homogeneous distribution (these two
\lq\lq opposite" cases share this second moment property)
\begin{equation}
\left< \rho^2(\vec x, t)\right> = \rho^2(t). 
\label{homogeneous}
\end{equation}

In any case, the balance of the three terms in
Eq.~(\ref{averagesum}) determines the behavior of the global density.
A general point that we return to subsequently in the context of particular
cases can already be made here: the presence of the \lq\lq driver"
$G(t)=\left<\gamma^2\right>$ causes deviations from classical behavior;
as already mentioned, a homogeneous
distribution of reactants at time $t$ is associated with a vanishing
$G(t)$ and with classical kinetics.

The behavior of $G(t)$ for {\em all times} and, in particular, how
rapidly it vanishes at long
times, is completely determined by the {\em initial}
configuration of reactants.  It is convenient to write
\begin{equation}
G(t) = \frac {1}{V} \sum_{\vec{k}} S(\vec{k}) e^{-2Dtk^2}
\label{explicitsum}
\end{equation}
or, in the large volume limit, 
\begin{equation}
G(t) = \frac {1}{(2\pi)^d} \int d\vec{k} S(\vec{k}) e^{-2Dtk^2},
\label{explicit}
\end{equation}
where we have used Eq.~(\ref{sln2}).
In either case $S(\vec{k})$ is the Fourier transform of the initial
correlation function,
\begin{equation}
S(\vec{k}) = \int d\vec{x} e^{-i\vec{k} \cdot \vec{x}}
\left< \gamma(\vec{x},0) \gamma(\vec{0},0)\right>.
\label{structure}
\end{equation}
$S(\vec{k})$ is called the structure function of the initial
configuration.  For $t>t_a$, where
\begin{equation}
t_a\equiv (Dk_a^2)^{-1},
\label{tc}
\end{equation}
$G(t)$ is essentially determined by the behavior of the
structure function for $k<k_a$.
In other words, the long-time behavior of the segregation function
defined by the difference variable
is dictated by the long-wavelength components of the initial distribution
(and also note that this behavior is entirely determined
by diffusion and is not influenced by the reaction).
In this paper we consider a variety of initial conditions characterized by
different forms of the structure function.  Familiar cases include the
random initial distribution in an infinite volume, for which
$S(\vec{k})=const$ for $k>0$, and
a random initial distribution of $A-B$ pairs separated by a distance $c$ in
an infinite volume,
for which $S(\vec{k})\sim k^2$ for $k< c^{-1}$. We return to these cases
subsequently. In general, if $S(\vec{k})\sim k^{\alpha}$ for $k<k_a$, then
a rescaling of variables in Eq.~(\ref{explicit}) immediately shows that
for $t>t_a$
\begin{equation}
G(t)\sim t^{-\frac{\alpha+d}{2}}.
\label{asymptotic}
\end{equation}
Note that $\alpha$ may be positive or negative (or zero). 
If $\alpha$ is sufficiently large (indicating a drastic cutoff of
the long wavelength components), specifically if $\alpha>4-d$, then
$G(t)$ decays more rapidly than $t^{-2}$ and it rapidly ceases to
play a role in the kinetics. The behavior then becomes
classical
[Vitukhnovsky {\em et al.}, 1988; Sancho {\em et al.}, 1996].
Conversely, any initial distribution that emphasizes the long
wavelength components in the density difference leads to a slower decay of
the \lq\lq driver" ($\alpha < 0$) and hence to greater deviations from
classical behavior.  We present examples of both cases.
Note that it would in any case at this point be premature to conclude
from Eq.~(\ref{asymptotic}) that $\rho(t)$ decays as 
$t^{-(\alpha+d)/4}$ -- at the very least, the moment relation
(\ref{homogeneous}) must be established first.

Ultimately we are constrained by the finite systems in which
our numerical work is carried out; this means that eventually $G(t)$
vanishes exponentially, reflecting the fact that there are no
contributions to it whose wavelengths are longer than the size of the
system.

The results of our analysis of the reaction-diffusion equations will be
compared with the numerical integration of Eq.~(\ref{difreac}). This
integration is carried out by discretizing the volume $V=L$  into
$N$ cells of size $\Delta x$ ($L=N\Delta x$) and time into
intervals $\Delta t$.  The Laplace operator is represented in the
standard symmetrical form with respect to the mesh size $\Delta x$,
and the time integration follows a standard first order Euler algorithm.
Note that adjustment of the discretization parameters $\Delta x$ and
$\Delta t$ can alternatively be accomplished by adjusting the parameters
$D$, $K$, and $N$.  We always set $K\gg D$ in order to be in the
diffusion-limited regime for a meaningful interval of time,
but otherwise set these parameters so as to
be able to clearly illustrate the behaviors under consideration.
Throughout we set $\Delta x = 1$, and $\Delta t = 0.01$
except for early time calculations where on occasion we use smaller values
of $\Delta t$ for the first few hundred integration steps.
Initially an amount $\rho_0$ of reactant $A$ is placed in some cells and
not in others, according to a specified distribution.  Similarly for
reactant $B$. If each reactant is initially placed in $n\leq N$ cells, then
the initial global density is $\rho_{init}=n\rho_0/N$.  Note that the maximum
initial density possible is then $\rho_{max}=\rho_0$, which occurs if every
cell is initially occupied by both reactants.
An average over different initial conditions is taken if necessary.

\section{A Variety of Initial Conditions }
\label{initial}

The initial pattern of fluctuations determines for all time the form of the
driver $G(t)$ in Eq.~(\ref{averagesum}) with (\ref{driver}).  We establish
below through specific examples that this driver decays more slowly
(rapidly) when the long wavelength components of the initial difference
fluctuations are enhanced (suppressed), and that the kinetics of the
$A+B\rightarrow 0$ reaction are more anomalous, i.e., further from
classical behavior, when the driver decays more slowly.  
When the decay of $G(t)$ is more rapid than $t^{-2}$ then its contribution
becomes irrelevant and the rate law approaches the classical form.

In this section we present the kinetic progressions for the $A+B\rightarrow
0$ reaction for a number of initial conditions.  The case of the random
initial condition has been extensively studied in the literature and is
reviewed here in detail because it serves as a benchmark for the
discussion of other initial conditions
[Argyrakis {\em et al.}, 1993; Lindenberg {\em et al.}, 1996].
This review allows us to establish some
limitations imposed by the reaction-diffusion approach
[Lindenberg {\em et al.}, 1996].
We then present an example of
an initial condition in which long wavelength components of the
density difference fluctuations are enhanced relative to those of a random
distribution.  This initial condition, in the form of a fractal pattern,
leads to more pronounced deviations from classical behavior.  We also
present an example of an initial distribution whose long wavelength
components are more constrained than the random, namely, the case of
correlated pairs [Lindenberg {\em et al.}, 1994, 1996, Sancho {\em et
al.}, 1996].

\subsection{Random initial distribution}

The random initial distribution of reactants has been the most thoroughly
studied, even though it is difficult to implement in practice.
Nevertheless, it is important because it is the distribution
that was first used to quantify the fact that diffusion in low dimensions
is not an effective mixing mechanism.  It was the distribution that
clearly showed that initial (microscopic) fluctuations are a
necessary element to achieve (macroscopic) segregated patterns, that is,
that initial inhomogeneities, with no further explicit separating
mechanism or interactions other than the reaction itself, lead to
macroscopic species segregation in low dimensions 
[Ovchinnikov \& Zeldovich, 1978; Toussaint \& Wilzcek, 1983;
Kang \& Redner, 1984; Kopelman, 1988; Burlatskii {\em et al.}, 1989].

We have argued elsewhere
[Argyrakis {\em et al.}, 1993; Lindenberg {\em et al.}, 1996]
that at extremely early times the kinetics of the
system is classical, that is, that with a random distribution of reactants
the global density obeys the classical rate law $d\rho(t)/dt =-K
\rho^2(t)$.  We also noted that this behavior in low dimensions
persists only for a time of order $K^{-1}$; on this time scale
the distribution already begins to deviate from the random distribution
because of the reaction of physically contiguous $A$ and $B$ molecules.
Because here we compare our analytic results with those obtained numerically
according to the discretization scheme described earlier, we are constrained
from consideration of initial concentration fluctuations on a spatial
scale smaller than $\Delta x$. In our discretized reaction-diffusion approach 
$A$ and $B$ densities that coexist within one cell are consumed on the
time scale $K^{-1}$, and each cell is, after a time $K^{-1}$,
either empty or occupied
essentially only by $A$ or only by $B$
[Sancho {\em et al.}, 1996].

\begin{figure}[htb]
\begin{center}
\leavevmode
\epsfxsize = 2.0in
\epsffile{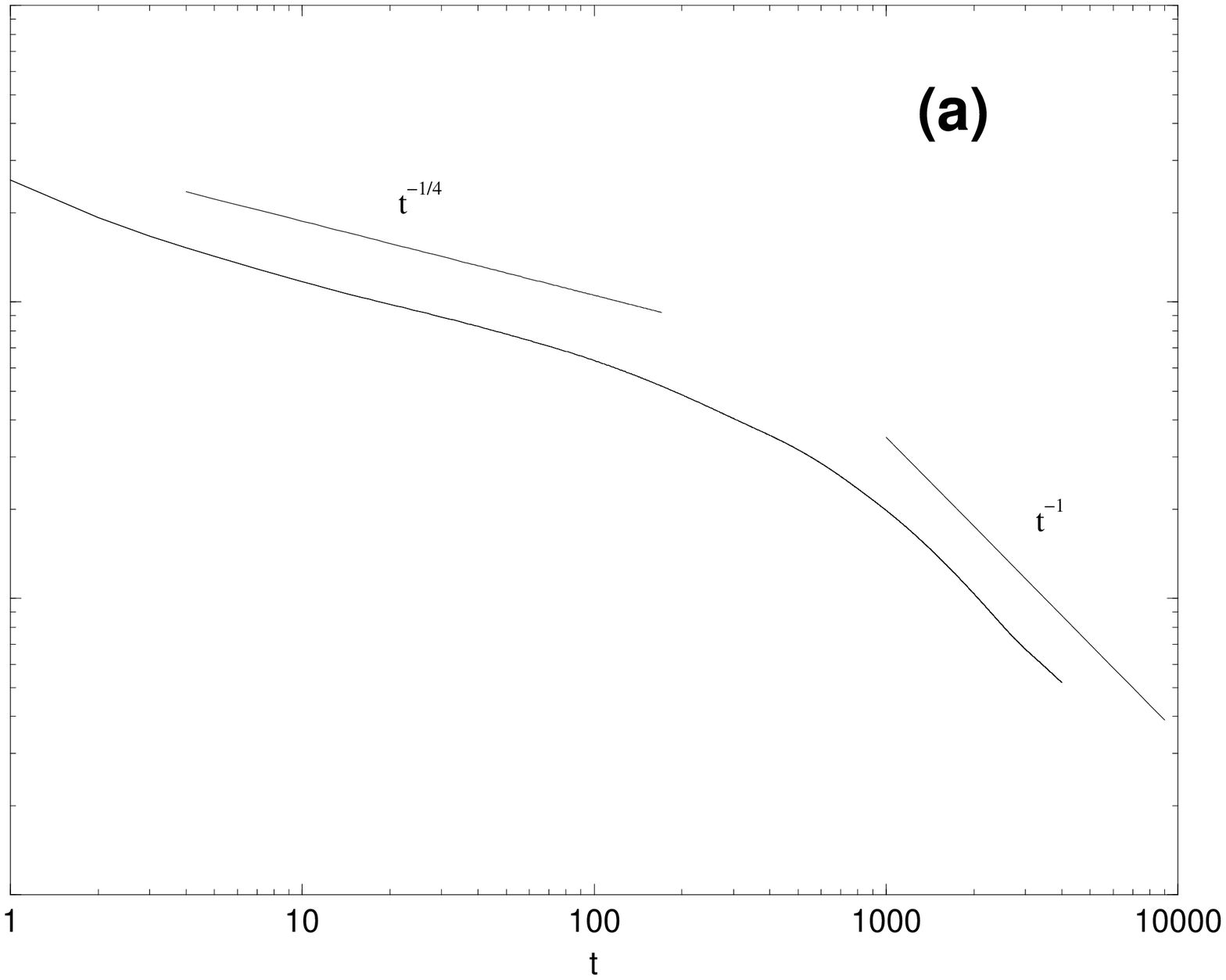}
\vspace{.1in}
\leavevmode
\epsfxsize = 2.0in
\epsffile{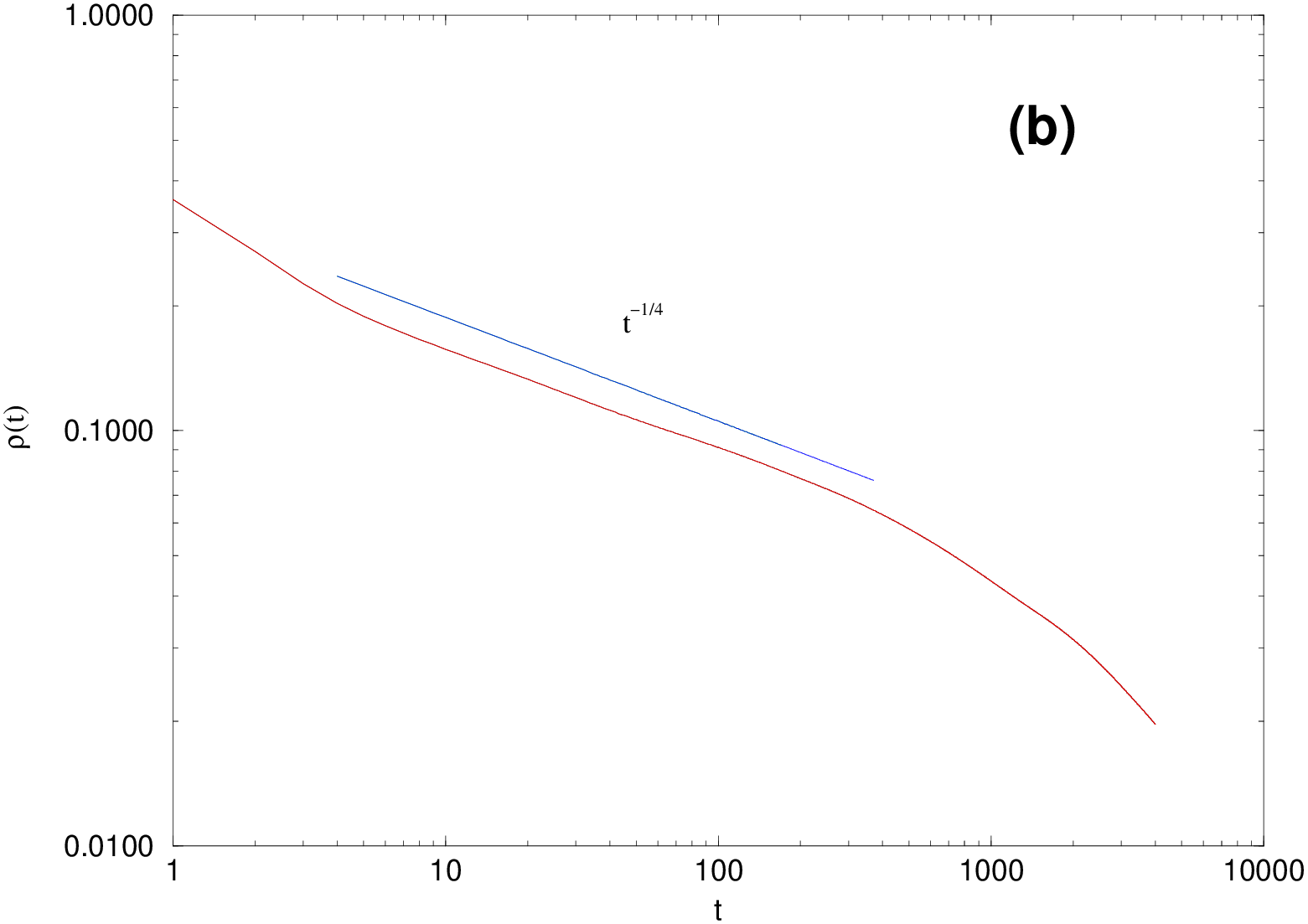}
\end{center}
\caption{
Decay of the global density in one dimension
from an initially random distribution of reactants.  In both figures
$N=512$
and $K=10$.  (a) $D=2.0$, (b) $D=0.5$. The curves are the numerical
solutions of the reaction-diffusion equations. Straight lines in (a):
$t^{-1/4}$ and $t^{-1}$. Straight line in (b): $t^{-1/4}$.}
\label{fig1}
\vspace{0.2in}
\end{figure}

Beyond this very early time behavior, the results that we detail
below are captured in Fig.~\ref{fig1}.  We first describe the expected
behavior and then compare it to that seen in the figure.
As noted earlier, an initially random distribution in
an infinite volume is characterized by the structure function
$S(\vec{k})=const$ for
$k>0$.  In our discretized integrations over a finite
volume, a random distribution is characterized by a constant structure
function only in the range $k_L<k<k_0$, where $k_0^{-1}=\Delta x$ is
the limit imposed by the discretization and $k_L$ is the limit imposed
by the finite system size.  It follows from Eq.~(\ref{explicit}) that
for times $t> t_0$, where
\begin{equation}
t_0=\frac{(\Delta x)^2}{D},
\label{earlytime}
\end{equation}
the segregation function $G(t)$ decays as $t^{-d/2}$
[cf. Eq.~(\ref{asymptotic})]; in fact, exact
integration of Eq.~(\ref{explicit}) leads to
(we return to finite volume effects in a moment)
\begin{equation}
G(t) = Q (Dt)^{-d/2},
\label{usual}
\end{equation}
where $Q$ is a constant proportional to the initial density.

The consequent behavior of the global density predicted by
Eq.~(\ref{averagesum}) then depends on the relation between the average
of the square of the density and the global density itself.  We have
noted that for a
random distribution Eq.~(\ref{random}) holds.  This relation is nearly
satisfied for a nearly random distribution; that is,
at early times before the
reactant distribution deviates too much from randomness the
contribution linear in the density dominates and the rate law is of
the form
\begin{equation}
\label{randomrate}
\frac{d\rho(t)}{dt}\sim -K[\rho_{max}\rho(t) - Q(Dt)^{-d/2}].
\end{equation}
For $d=1$ the dominant terms are those on the right hand
side, and the global density thus decays as 
\begin{equation}
\rho(t)\ \sim t^{-1/2}.
\label{decay1}
\end{equation}
This decay is characteristic of the presence of a \lq\lq depletion zone"
around each reactant (and, for the same reasons, is also characteristic
for the $A+A\rightarrow 0$ reaction).
The critical dimension where all three terms are of the same order is
$d=2$.  Note that consistency in one dimension requires that 
the difference between the two contributions on the right decay in time
as $t^{-3/2}$ so as to be balanced by $\dot\rho(t)$.

Deviations from
randomness and, in particular, spatial segregation of reactants, begin
to set in very quickly for $d< 4$ (for a detailed discussion of
crossover times see
[Argyrakis {\em et al.}, 1993; Lindenberg {\em et al.}, 1996]).
The moment relation changes to (\ref{homogeneous}), and the
rate law becomes
\begin{equation}
\label{Zeldovichrate}
\frac{d\rho(t)}{dt}\sim -K[\rho^2(t) - Qt^{-d/2}].
\end{equation}
For $d<4$ the dominant terms are again those on the right hand
side, and the global density in one dimension now decays as 
\begin{equation}
\rho((t)\ \sim t^{-1/4}
\label{decay2}
\end{equation}
(\lq\lq Zeldovich" behavior).
The critical dimension where all three terms are of the same order is
$d=4$.  In one dimension the difference between the dominant terms,
and also $\dot \rho(t)$, now decay as $t^{-5/4}$.
The Zeldovich behavior reflects the well-known segregation of the
species -- even as the densities decrease, the sizes of the regions
essentially occupied by only one species increase.  Note that the fact
that (\ref{homogeneous}) rather than (\ref{random}) holds
in this segregated regime indicates that
the spatial distribution of the reactants {\em within} the single-species
aggregates is not random.  If it were random, a relation such as
(\ref{random}) (with slightly modified coefficients) would hold instead.

Eventually, when $t \rightarrow t_f$ [cf. Eq.~(\ref{finitesizet})], the
above analysis must be adjusted to reflect finite system size effects.  The
integral form Eq.~(\ref{explicit}) can no longer be used, and one must
return to the sum form Eq.~(\ref{explicitsum}).  The segregation
function for
$t\gg t_f$ decays exponentially.  There is
a rapid drop in the global density due to the
increased reaction rate caused by the mixing of species forced
by the boundaries of the system.  A \lq\lq homogeneous" regime is
then reached where the
last term in Eq.~(\ref{averagesum}) is negligible, and the rate law
returns to the classical form
\begin{equation}
\label{classicalend}
\frac{d\rho(t)}{dt}\sim -K\rho^2(t), 
\end{equation}
with the associated decay
\begin{equation}
\rho(t) \sim (Kt)^{-1}.
\label{decay3}
\end{equation}

Figure~\ref{fig1} shows the evolution of the global density for a
one-dimensional system with a random initial distribution of cells
containing $A$ and a random initial distribution of cells containing $B$. 
We generate each of these distributions separately
but with the same number of occupied cells for each species,
and then simply
superpose them. This produces some \lq\lq premixed" cells whose
density decays very rapidly
[Sancho {\em et al.}, 1996].
Extensive
premixing is avoided by starting with a relatively low initial
density. We calculate the resulting evolution
for two different values of the diffusion
coefficient.  Figure~\ref{fig1}(a) is for a larger diffusion coefficient
than Fig.~\ref{fig1}(b).  In both, the Zeldovich decay regime is clearly
seen over a considerable portion of the figure -- the numerical
integration results run parallel to the $t^{-1/4}$ lines. 
The earliest time depletion zone behavior that precedes the Zeldovich
regime, where $\rho(t)$ is expected to decay
as $t^{-1/2}$, is more apparent in the system with the smaller diffusion
coefficient. We have shown elsewhere that the transition from depletion zone
behavior, $t^{-1/2}$, to Zeldovich behavior, $t^{-1/4}$, occurs at a time
$t_s$ inversely proportional to the diffusion coefficient.
The rapid decrease in the density
due to the onset of finite size effects is clearly visible in both cases.
For the larger diffusion coefficient (more rapid mixing) the onset of the
classical regime is seen clearly, as indicated by the fact that the density
decay curve in (a) runs parallel to the $t^{-1}$ line.
The time of the run
in Fig.~\ref{fig1}(b) is not sufficiently long 
to see this classical regime clearly.
Note that, as expected, the finite size effects set in later for
the system with the smaller diffusion coefficient. 

A point to note concerns the density moments
of a strictly random mixture of $A$'s and $B$'s and those of a homogeneous
mixture.  In both of these
the average of the product of the densities of different species is simply
the product of the averages: $\left< \rho_A\rho_B\right> =
\left<\rho_A\right>\left<\rho_B\right>$.  Consequently, an average of 
Eq.~(\ref{difreac}) immediately leads to second order kinetics in both cases.
It is interesting to note, however, that this equality of moments does not
extend to other moments; in particular, the sum and difference variables
for these two distributions behave quite differently. In a random
distribution there are fluctuations in both the total ($A$ plus $B$) local
density and also in the difference variable, while in a homogeneous
distribution these fluctuations have been smoothed out -- indeed, this
is what is meant by \lq\lq homogeneous." For a random
distribution the average of the square of the difference variable is
linear in the average density, $\left<\gamma^2(\vec x)\right>=
\rho_{max}\rho$, and the average of the
square of the sum variable also has such a linear contribution, cf.
Eq.~(\ref{random})
[Argyrakis {\em et al.}, 1993; Lindenberg {\em et al.}, 1996].
On the other hand, for a homogeneous distribution the
difference variable vanishes so that
$\left<\gamma^2(\vec x)\right>=0$,
while the average
of the square of the sum variable is quadratic in the average density,
cf. Eq.~(\ref{homogeneous}).  The kinetics in the reaction-diffusion picture
depend only on the difference
$\left<\rho^2\right> - \left<\gamma^2\right>$, which is the same for both
cases.

\subsection{A detour}

The above description of the evolution of the reaction-diffusion system
from an initial random distribution of reactants predicts the behavior
of the global
density in various kinetic regimes, Eq.~(\ref{decay1}) followed by
Eq.~(\ref{decay2}) followed by Eq.~(\ref{decay3}), and in some cases
predicts the
approximate time around which a transition from one kinetic behavior to
another takes place. However, we are not easily able to predict the
actual crossover behavior itself.  Thus, for example, we know {\em when}
finite size effects set in, and we can reason that the global density will
drop sharply around that time, but we have not predicted the functional
form of the drop.  

Some insights can be obtained by pausing at this point to analyze the
behavior of the different terms in the reaction-diffusion equation itself
in some detail.  This \lq\lq generic" examination will help us understand
not only the case of random initial conditions but also the behavior of the
system with other initial conditions.

\begin{figure}[htb]
\begin{center}
\leavevmode
\epsfxsize = 3.2in
\epsffile{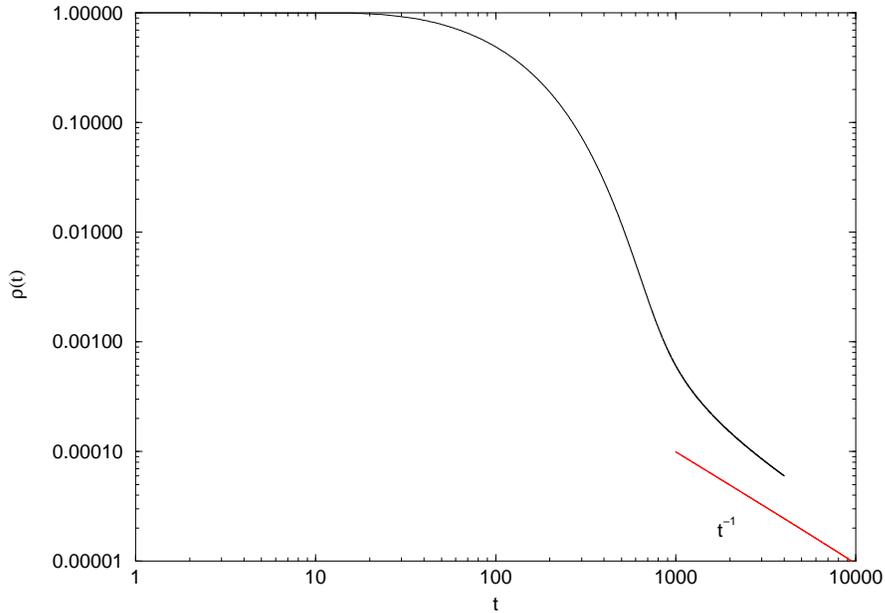}
\end{center}
\caption{
Decay of the global density in one dimension from point sources.
$N=64$, $D=2.0$, $K=10$, and the initial source cells are separated
by 31 empty cells.  The straight line is $t^{-1}$.}
\label{fig2}
\vspace{0.2in}
\end{figure}

The analysis is facilitated by considering the evolution of the global
density from the following initial condition (more easily achievable
than the random): $A$ is initially placed in a single cell centered at
$x_A$ and nowhere else, and $B$ is initially placed in a different single
cell centered at $x_B$ and nowhere else.  In our (one-dimensional)
numerical implementation of a system with $N=64$ cells we place these two
cells as far from one another
as possible, i.e., there are 31 empty cells between the
two initially occupied cells. The resulting evolution is shown in
Fig.~\ref{fig2}.  The global density essentially does not decay at all until
$A$'s and $B$'s on average encounter one another. 
In this numerical implementation the
time for these encounters is roughly the
same as the time at which finite size effects set in. There is
an abrupt decrease in the density and the system then
settles into classical behavior, as indicated by the straight $t^{-1}$
line. This system is fairly trivial
but offers an opportunity to make a number of observations about the
various terms in the reaction-diffusion equation. 

The segregation function for this initial condition in which segregation is
imposed from the outset is easily calculated in the large $L$ limit. In
one dimension we find
\begin{equation}
G(t)=\frac{Q}{\sqrt {Dt}}\left( 1 - e^{-(x_A-x_B)^2/8Dt}\right).
\label{pointdiff}
\end{equation}
As long as $t\ll t_m\equiv (x_A-x_B)^2/8D$, the segregation function
has the same time dependence, $t^{-1/2}$, as that of the random initial 
condition, viz. Eq.~(\ref{usual}) --
let us for the moment concentrate on this time regime.
As in the random case, the decay of the driver $G(t)$ is sufficiently
slow to be a dominant contribution in Eq.~(\ref{averagesum}).  As in the
random case, this dominant contribution must be balanced by the other
dominant contribution, $\left< \rho^2\right>$.  However there is a major
difference between this case and the random one: in the case of a random
initial condition $\left<\rho^2\right>$ is related in a simple
fashion to the global density $\rho(t)$,
cf. Eqs.~(\ref{random}) and (\ref{homogeneous}).  Therefore the balance
of the two terms on the right of Eq.~(\ref{averagesum}) immediately tells us
the kinetic behavior of the global density, as described
earlier.  That led to the depletion zone behavior at early
times ($t^{-1/2}$) followed by the Zeldovich behavior at later times
($t^{-1/4}$). The balance between $\dot\rho(t)$ and the difference
between $\left<\rho^2\right>$ and $\left<\gamma^2\right>$ simply confirms
this kinetic behavior.

In the present example there are two (related) differences from the
random initial condition case.  First, there is {\em no} simple relation
between the average of the square of the sum variable and the global
density. Indeed, for times $t\ll t_m$ the reaction plays essentially no
role, so that one can quite accurately evaluate $\left<\rho^2\right>$
just from the
diffusion portion of the reaction-diffusion equation to obtain
\begin{equation}
\left<\rho^2(x,t)\right>\approx
\frac{Q}{\sqrt {Dt}}\left( 1 + e^{-(x_A-x_B)^2/8Dt}\right).
\label{pointsum}
\end{equation}
On the other hand, the global density is constant in this time regime,
so that there
is no simple moment relation between $\left<\rho^2\right>$ and the global
density.  The leading contributions of Eqs.~(\ref{pointdiff})
and (\ref{pointsum}) cancel as they
do in the random case, but the difference between them is now exponentially
small.  Here this in itself says nothing about the decay of the
global density, since this difference is not related in any simple way to
the global density. Information about the global density is obtained
only when we explicitly balance $\dot \rho (t)$ against this
very small difference.  

In our numerical example the time $t_m$ when reactants encounter one
another on average is of the same order as the time $t_f$. At around
this time the
segregation function $G(t)$ drops abruptly.  So does the global density,
which then settles to the classical decay law $(Kt)^{-1}$.  When this
happens, the segregation function becomes irrelevant, the moment relation
Eq.~(\ref{homogeneous}) becomes applicable as diffusion homogenizes the
system, and the principal balance in Eq.~(\ref{averagesum}) is between
$\dot\rho(t)$ and the first term on the right, that is,
Eq.~(\ref{averagesum}) reduces to the classical rate law.

\begin{figure}[htb]
\begin{center}
\leavevmode
\epsfxsize = 3.2in
\epsffile{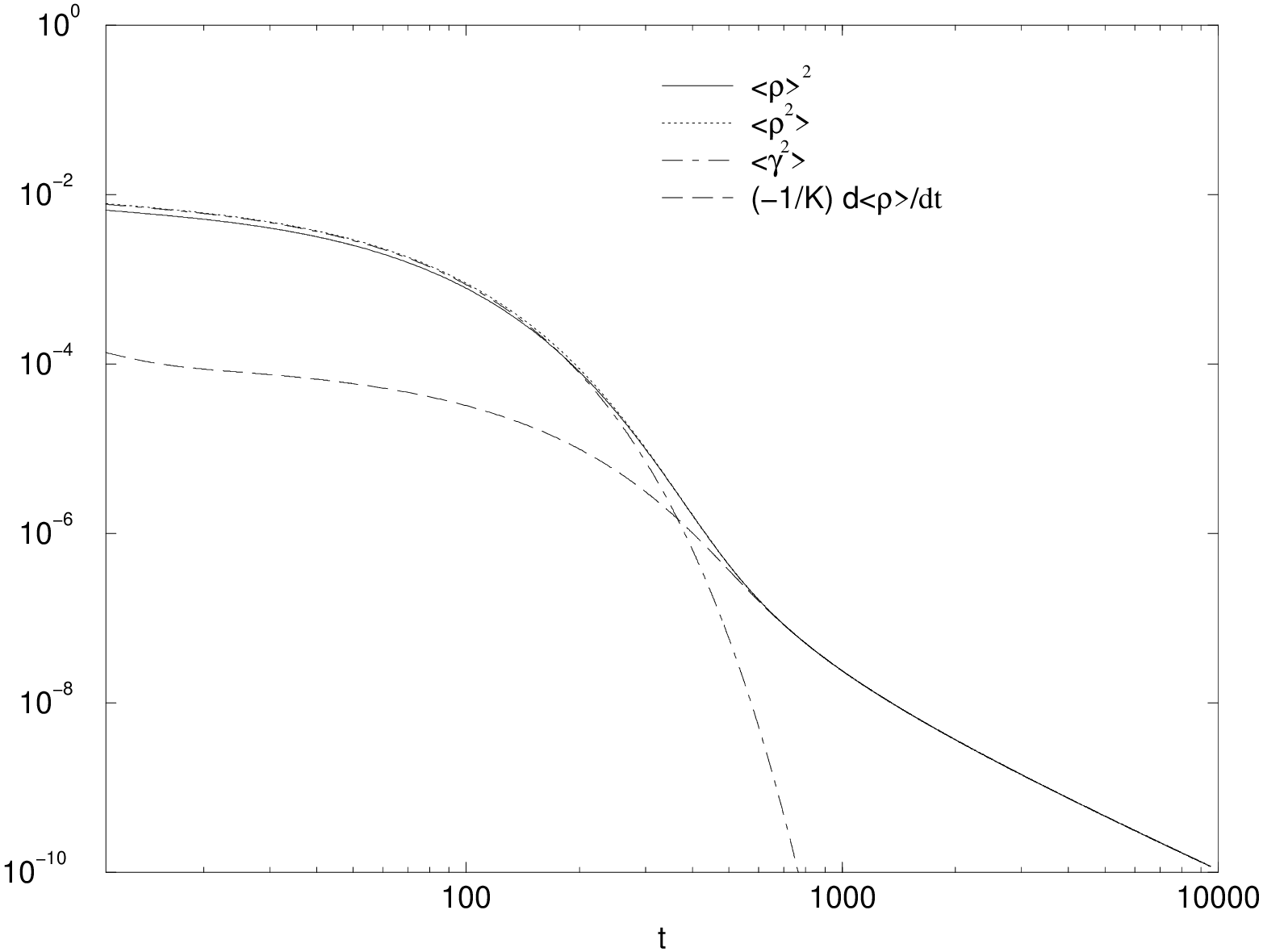}
\end{center}
\caption{
Schematic of the typical behavior of the various contributions to
Eq.~(\protect\ref{averagesum}\protect):
$G(t)=\left<\gamma^2\right>$,
$\left<\rho^2\right>$, $\rho^2$, $-\dot\rho /K$.}
\label{fig3}
\vspace{0.2in}
\end{figure}

This description together with that of the random initial condition case
permits us to reach some generic conclusions about the behavior of the
different contributions to Eq.~(\ref{averagesum}).
Figure~\ref{fig3} is a schematic (not to any particular scale)
of the evolution of these different
contributions in low dimensions.  First consider the segregation
function $G(t)=\left<\gamma^2\right>$ (solid curve).  It typically
begins by decaying according to a power law, more
slowly than $t^{-2}$, with perhaps a transition from one slow power decay
to another.  At some time, at the latest the time
$t_f$ when finite system size
effects set in, the segregation function drops abruptly (exponentially) as
the system becomes homogenized.  Next consider the average of the square
of the density, $\left<\rho^2\right>$ (dashed curve).  It is larger
than the segregation
function but decays to leading order in the same fashion before time $t_f$.
These two contributions are the leading terms in Eq.~(\ref{averagesum}) and
therefore balance one another. The difference between them
decays more rapidly 
than either term, so in this regime $\dot \rho(t)$ (which balances this
difference and is sketched as a dot-dashed curve) decays rapidly as
well and is negligible in the leading balance.
The square of the global density, $\rho^2(t)$, is also 
sketched (dotted curve).
It is generally smaller than the other two contributions, but approaches
$\left<\rho^2\right>$ (for different reasons in different cases).  As it
does so, the global density itself is determined by the leading balance,
that is, $\rho(t) \sim G^{1/2}(t)$, and this relation must of course be
consistent with the balance between $\dot \rho$ and the difference between
the leading terms.
During the crossover time when G(t) drops abruptly the system homogenizes,
and $\dot \rho$ catches up with $\rho^2$, which by then is essentially
equal to $\left<\rho^2\right>$.  $G(t)$ then becomes irrelevant and
the kinetics is classical.

This generic sequence describes every initial condition that we have dealt
with here and elsewhere
[Argyrakis {\em et al.}, 1993; Lindenberg {\em et al.}, 1994, 1996;
Sancho {\em et al.}, 1996].
The differences arise in the
specific slopes associated with a variety of anomalous kinetics
(that is, decays of the global density that are slower than the classical
$t^{-1}$) before finite size effects set in.  These anomalous kinetics
arise from (indeed, require) initial density fluctuations and
show a range of behavior depending on the distribution of these
fluctuations. The random initial condition is but one example, and in one
dimension is associated with decays of the global density of the
form $t^{-1/2}$ followed by $t^{-1/4}$.  In the examples to follow we deal with
the \lq\lq more anomalous case" of a fractal initial distribution leading
to a slower decay than that of the random case, and with a \lq\lq less
anomalous" case of initially correlated pairs leading to a decay closer to
the classical.

\subsection{Fractal initial distribution}

Consider next an initial condition in which the long wavelength components
of the difference variable are emphasized relative to the random initial
distribution, leading us to expect a greater deviation from classical
behavior
[Provata {\em et al.}, 1996].

Our initial distribution is generated as follows: the number
$N$ of cells in a row is chosen to be a power of $3$, $N=3^n$.
We divide $N$ by 3, thereby generating $N/3$ domains (each containing many
cells).  One of the three
domains, chosen at random from among the three, remains empty and is
discarded from further consideration.  The other
two domains are each subdivided into three subdomains, and one of
each group of three subdomains again remains empty and is set aside, while
the other two are again subdivided. 
The process continues in this way until the selected domains are as small
as possible, that is, of size $\Delta x$. 
An initial density $\rho_0$ of one of the reactants is placed in
each of the selected elementary cells. The process is repeated for the
second reactant, and these two distributions are superimposed.
In the limit $N\rightarrow\infty$ this leads to a fractal geometry of
occupied cells of dimension $D_f = \ln2/\ln3 = 0.63$
[Vicsek, 1992].
The variable $\gamma$ is also characterized by the same initial
fractal dimension, and the
structure function for small $k$ behaves as a power law with an exponent
given by the fractal dimension,   
\begin{equation}
S(k) \sim k^{-D_f}
\label{fractal}
\end{equation}
Note that the exponent is negative, i.e., the long wavelength components
are {\em enhanced} relative to those of a random intitial distribution.

Consider now the sequence of behaviors that might be expected with this
initial condition.  Since there may be some premixed cells, there will
be a rapid decay of density in these cells on a time scale $K^{-1}$.
Once this is over, we expect a lull in the activity since aside from the
premixed cells, occupied cells are in general quite separate. 
Indeed, if
the fractal dimension is too high (e.g. if instead of subdividing into 3
parts at each generation we subdivided into $m$ portions with $m$ fairly
large) then there are many premixed cells and
much of the reactant disappears during the very early time regime. The
quiescent period then dominates the time evolution because there is little
reactant left and it takes a very long time for reactants to find one
another (somewhat like our point source example). 

During the quiescent period the distribution is rather segregated
but the sum variable is becoming spatially homogenized through
diffusion; hence one
expects the moment relation (\ref{homogeneous}) to apply.  The
difference variable, on the other hand, is determined by
the initial fractal pattern and leads to the segregation function
determined by Eq.~(\ref{fractal}). The rate is then of the form 
\begin{equation}
\label{fractalrate}
\frac{d\rho(t)}{dt}\sim -K[\rho^2(t) - Q(Dt)^{-(d-D_f)/2}].
\end{equation}
For $d=1$ (and, indeed, for $d < 4+D_f$) the dominant terms are those on
the right hand side, and the global density thus decays as
\begin{equation}
\rho((t)\ \sim t^{-(1-D_f)/4}.
\label{decay}
\end{equation}
Note that this decay is {\em even slower} than the Zeldovich behavior.
This decay continues until
finite size effects and the subsequent behavior described earlier
take over.

\begin{figure}
\begin{center}
\leavevmode
\epsfxsize = 2.0in
\epsffile{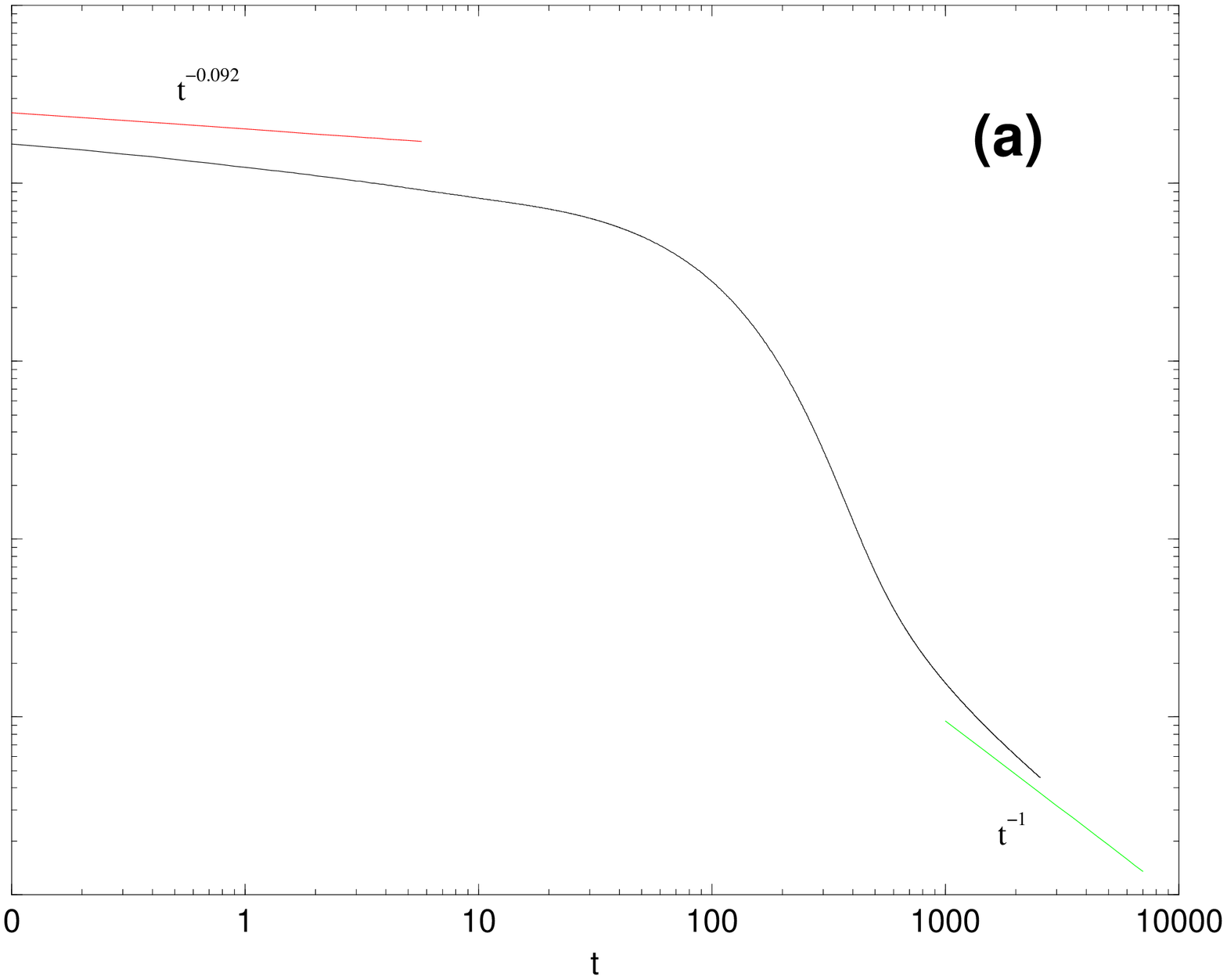}
\vspace{.1in}
\leavevmode
\epsfxsize = 2.0in
\epsffile{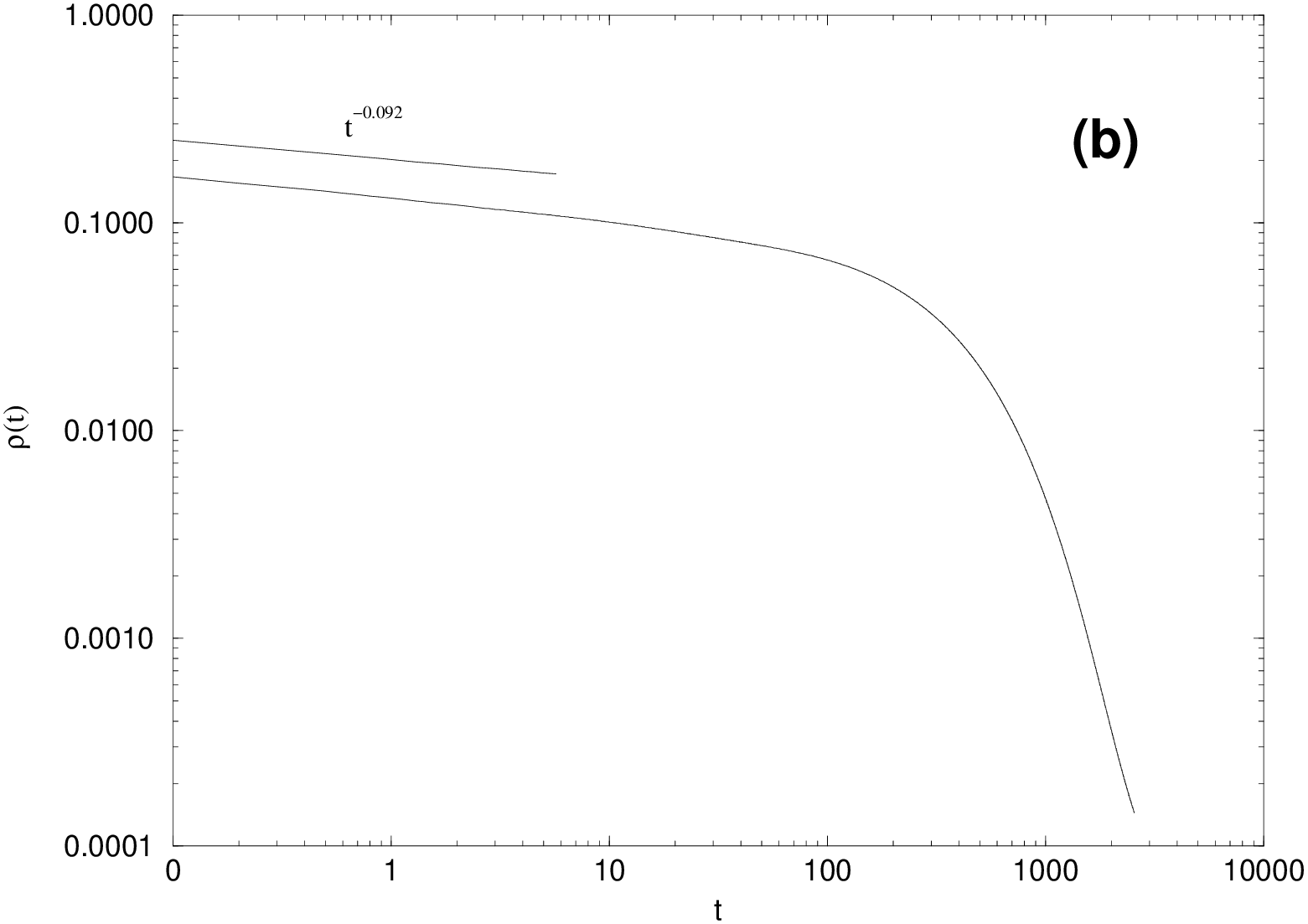}
\end{center}
\caption{
Kinetics for the fractal-like initial pattern described in the text.
In both figures $N=81$ and $K=10$.  (a) $D=2.0$, (b) $D=0.5$. The curves
are the numerical solutions of the reaction-diffusion equations. Straight
lines in (a): $t^{(1-D_f)/4}=t^{-0.092}$ and $t^{-1}$.  Straight line in
(b): $t^{-0.092}$.}
\label{fig4}
\vspace{0.2in}
\end{figure}

In Fig.~\ref{fig4} we show the results of the numerical integration of the
reaction-diffusion equations and the prediction (\ref{decay}). 
It should also be noted that our initial patterns are not
\lq \lq ideal" fractals because we stop the subdivision at a
finite cell size.  Our systems have $N=81$, so we only go through
four generations of subdivisions.  Nevertheless, the predicted results are
clearly apparent in the figures.  In Fig.~\ref{fig4}(a) we see the slow
decay of the form $t^{-0.092}$ expected for an initial distribution
of fractal dimension $D_f=0.63$, followed by the
abrupt decrease
in the global density associated with finite system size effects and the
eventual approach to classical behavior.  Figure~\ref{fig4}(b) is for a
smaller diffusion coefficient -- the slow decay continues for a longer
time, finite system size effects set in later, and the time scale is not
sufficiently long to see the eventual appearance of classical behavior.

\subsection{Initially correlated pairs}

A generic example of a constrained initial distribution of reactants in
which the long wavelength components of the initial density difference are
diminished through built-in correlations is that in which
randomly oriented $A-B$ pairs of particles 
separated by a fixed distance $c$ are initially distributed
randomly in the system.
Subsequent to this initial constrained placement, each particle moves
freely.  Lindenberg et al.
[Lindenberg {\em et al.}, 1994, 1996;
Sancho {\em et al.}, 1996]
have studied in detail the kinetic progression for the
$A+B\rightarrow 0$ reaction with this initial distribution.
In our discretized reaction-diffusion picture we translate this into
the initial random placement of pairs of cells with centers
separated by a distance $\lambda$ ($c=\lambda \Delta x$),
one cell of each pair containing only species $A$ and the other only
species $B$
(whether or not we allow initial occupancy of a cell by different species
arising from different pairs
is immaterial - it only changes the very early transient, as we have
noted above).

The segregation function for this initial condition in an infinite volume
is given by
[Lindenberg {\em et al.}, 1994, 1996]
\begin{equation}
G(t) = Q (Dt)^{-d/2}\left( 1-e^{-\lambda^2/8Dt}\right),
\label{couples}
\end{equation}
where, as before, $Q$ is a constant proportional to the initial density.
At times $t<t_c \equiv \lambda^2/8D$ the segregation function behaves
exactly as in the case of the random initial distribution.  Beyond time
$t_c$ the decay of $G(t)$ changes from $t^{-1/2}$ to $t^{-3/2}$ (the
associated behavior of the structure function for small $k$ is $S(k)\sim
k^2$)
[Sancho {\em et al.}, 1996].
This small $k$ behavior reflects the constraints
on the long wavelength components of the initial density difference
fluctuations imposed by the pair correlations.

The decay of the global density depends, as always, on the balance 
of terms in Eq.~(\ref{averagesum}) and on the relation between $\left<
\rho^2\right>$ and the global density. 

In order to describe the ensuing kinetic regimes, we must distinguish two
cases depending on the relation between the time $t_s$ and the time
$t_c$.  Recall from our discussion of the random initial condition
that $t_s$ is the crossover time at which the relation between the second
and first moment of an initially random distribution changes from
the form Eq.~(\ref{random}) to Eq.~(\ref{homogeneous}), that is, for
the segregation
process to begin.   If $t_c \leq t_s$ then segregation never sets in
because initially correlated pairs of cells react (in a \lq\lq burst")
before segregation can begin.  We have discussed this case in detail in
[Lindenberg {\em et al.}, 1994, 1996]
and will not present this particular case
here.  If $t_c > t_s$ then up to time $t_c$ the system evolves as would
a random distribution, with an early $\rho\sim t^{-1/2}$ depletion zone
evolution followed by a $\rho \sim t^{-1/4}$ Zeldovich evolution during
which segregated regions grow in size.   These two regimes are described by
Eqs.~(\ref{randomrate}) and (\ref{Zeldovichrate}) respectively.   However,
at around time $t_c$ the kinetic behavior changes because the second term
in Eq.~(\ref{couples}) becomes important: as $t$ increases well beyond
$t_c$, instead of Eq.~(\ref{Zeldovichrate}) we now have
for $d<2$ a rate law of the form
\begin{equation}
\label{shrink}
\frac{d\rho(t)}{dt}\sim -K[\rho^2(t) - Q(Dt)^{-(d+2)/2}].
\end{equation}
The global density in one dimension now decays as
\begin{equation}
\rho((t)\ \sim t^{-3/4}.
\label{decay4}
\end{equation}
We have explained the origin of this behavior (note that this decay,
although anomalous, is faster than that of an $A+A$ reaction): 
the distance $\lambda$ effectively sets a limit on the sizes of the
segregated aggregates. The initial correlations in effect homogenize the
system by constraining the fluctuations of wavelength greater than
$\lambda$.  The relatively rapid decay Eq.~(\ref{decay4}) occurs because
after time $t_c$ the single species aggregates
actually begin to shrink. It is this shrinkage that
is reflected in the more rapid decay. 

Eventually system size
effects enter the picture, Eq.~(\ref{couples}) can no longer be used, and
$G(t)$ decays, as always, exponentially.  The kinetic
evolution beyond this point is classical.

\begin{figure}[htb]
\begin{center}
\leavevmode
\epsfxsize = 3.2in
\epsffile{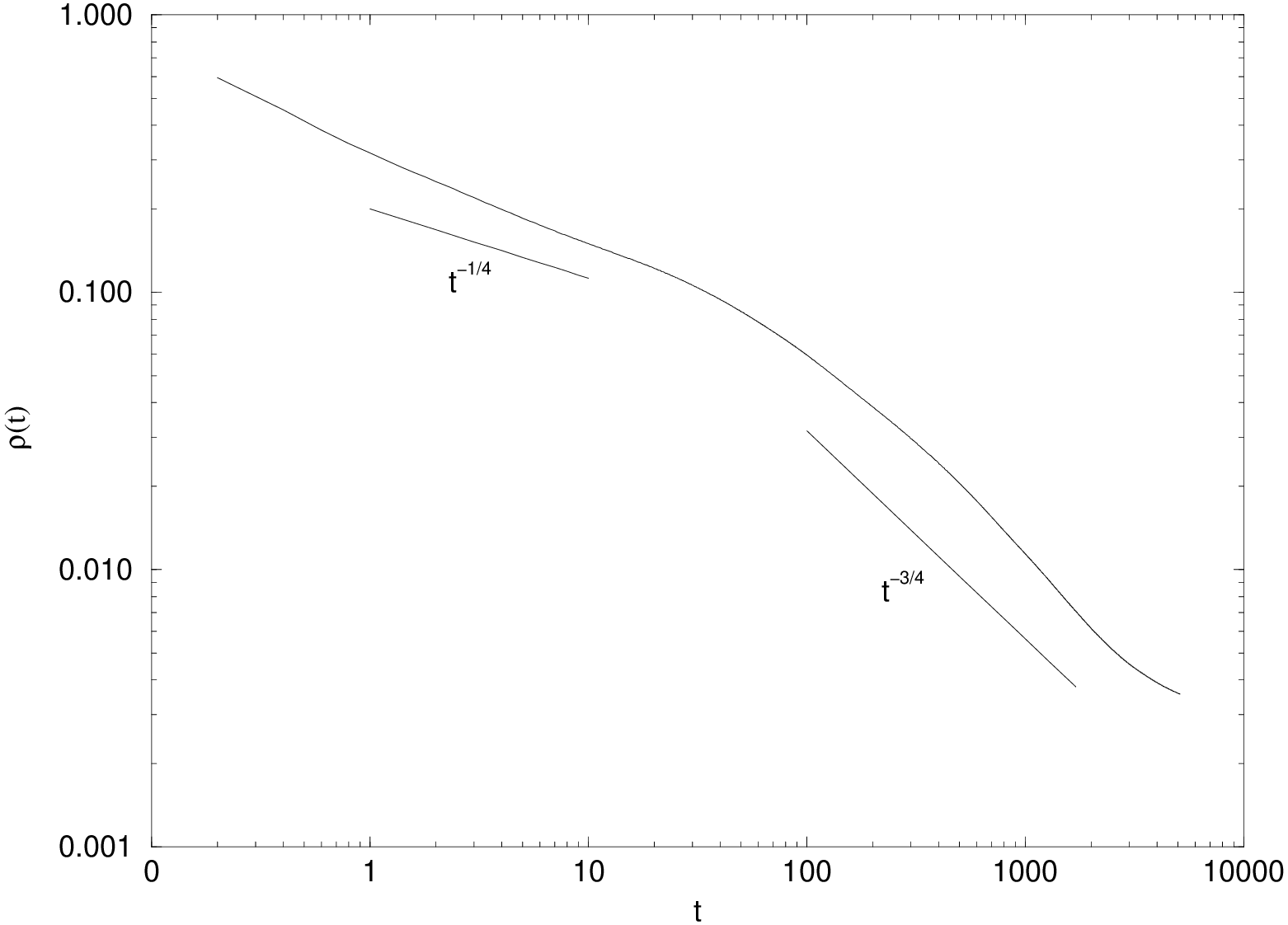}
\end{center}
\caption{
Kinetics for initially correlated pairs. $N=4096$, $K=10$, $D=2$, and
$\lambda =32$.
The curve is the numerical solution of the reaction-diffusion equations.
Straight lines: $t^{-1/4}$ and $t^{-3/4}$. }
\label{fig5}
\vspace{0.2in}
\end{figure}

Figure~\ref{fig5} illustrates the salient features of this behavior. 
We first see a decay of the global density close to the Zeldovich 
$t^{-1/4}$, followed by a crossover to a $t^{-3/4}$ aggregate
shrinkage decay beginning at around the time $t_c\sim \lambda^2/8D = 64$. 
Note that for the parameters used in this figure $t_f$ is off the
scale and finite size effects are thus not seen.

\section{Conclusions}
\label{conclusions}

We have presented a study of the effects of initial distributions on
the behavior of the irreversible reaction $A+B\rightarrow 0$ 
under stoichiometric conditions, mainly in one dimension ($d=1$). 
Our work is based on a reaction-diffusion model, and
we include detailed discussions of the behavior of each term in the
reaction-diffusion equations.
We have focused on the effects of the initial pattern of
fluctuations of the density difference and have stressed that this pattern
determines the evolution of the
system for all time. In particular, the long-wavelength
components of the initial difference distribution determine the asymptotic
decay of the reactant densities. Thus, this system constitutes
a temporal \lq\lq mirror" that for all time reflects the instantaneous
initial spatial pattern.  

An initially random distribution of
reactants leads to well-known and amply studied 
segregation of species and the associated slowing of the reaction.
In an infinite system the asymptotic decay of the reactants in this case
is described by the \lq\lq Zeldovich law" $\rho\sim
t^{-d/4}=t^{-1/4}$, which is distinctly different from the 
classical law-of-mass-action behavior $\rho\sim t^{-1}$.  Initial
distributions that limit the long wavelength components of the 
difference fluctuations lead to constraints on the macroscopic segregation
effects and to rate laws closer to the classical.  We reviewed a particular
example of such an initial distribution, namely, one containing initially
correlated $A-B$ pairs.
On the other hand, initial distributions that
enhance the long wavelength components lead to even greater
deviations from
the law of mass action.  For instance, an initial fractal difference
pattern of dimension $D_f$ leads to the very slow decay
$\rho\sim t^{-(d-D_f)/4}$.

We have emphasized the fact, also well known, that no single exponent
characterizes the density decay for all time, even in an infinite system. 
Instead, we have indicated
that it is often possible to characterize the decay by a single exponent
for a considerable length of time (compared to the other time scales in the
problem as determined by the reaction rate coefficient and the diffusion
coefficient), followed by a crossover of relatively short duration that
leads to another exponent for a considerable time span, etc. 
The sort of analysis presented here is only appropriate when these
time scale separations are possible. 

Our numerical work is based on a direct integration of the
reaction-diffusion equations and is, in all cases, subject to the long
wavelength constraints imposed by a finite system size.  Thus, although the
anomalous decays indicated above might be expected to persist forever 
for irreversible diffusion-limited reactions in
infinite systems, in the present work we do not emphasize the idealized
asymptotic behavior. Instead, we include a discussion
of the behavior beyond the time $t_f$ at which finite
system size effects become apparent. The reaction-diffusion model predicts
classical behavior at long times. We should note that although in the paper
we discuss averages over initial distributions, the numerical
results presented in our figures are all for single realizations of these
distributions.  Further averaging over initial distributions did not modify
the outcomes.

It would be interesting to observe in detail not only the time evolution of
the reactant densities but also the spatial patterns associated with this
evolution for the different initial distributions that we have studied.  We
plan to do so in the future.

\section*{ACKNOWLEDGMENTS}

J.M.S acknowledges the kind hospitality of the Institute for Nonlinear
Science and of the Department of Chemistry and Biochemistry at the
University of California, San Diego where this work was initiated.
We gratefully acknowledge financial support of the U.S. Department of
Energy Grant No. DE-FG03-86ER13606 and of the Direcci\'{o}n General de
Investigaci\'{o}n Cient\'{\i}fica y T\'{e}cnica (Spain) under
Project No. PB93-0769.





\end{spacing}

\end{document}